\documentclass[11pt]{article}
\usepackage{amsmath}
\usepackage{amsfonts}
\newcounter{mnotecount}[section]

\newcommand{\koniec}{\begin{flushright}  $\Box $ \end{flushright}}
\newtheorem{theo}{Theorem}[section] 
\newtheorem{prop}[theo]{Proposition}  

\newtheorem{defi}[theo]{Definition}
\newtheorem{col}[theo]{Corollary}

\renewcommand{\themnotecount}{\thesection.\arabic{mnotecount}}

\newcommand{\mnote}[1]
{\protect{\stepcounter{mnotecount}}$^{\mbox{\footnotesize
$
\bullet$\themnotecount}}$ \marginpar{
\raggedright\tiny\em
$\!\!\!\!\!\!\,\bullet$\themnotecount: #1} }

\def\p{\partial}
\def\be{\begin{equation}}
\def\ee{\end{equation}}
\def\bea{\begin{eqnarray}}
\def\eea{\end{eqnarray}}

\def\ov{\overline}

\newcommand{\spp}{\mathbb{S}}
\newcommand{\CP}{\mathbb{CP}}
\newcommand{\C}{\mathbb{C}}
\newcommand{\HH}{\mathbb{H}}
\newcommand{\R}{\mathbb{R}}

\newcommand{\hook}{{\setlength{\unitlength}{11pt}   
                   \begin{picture}(.833,.8)
                   \put(.15,.08){\line(1,0){.35}}
                   \put(.5,.08){\line(0,1){.5}}
                   \end{picture}}}

\begin{document}

\title{\vskip -70pt
\begin{flushright}
{\normalsize DAMTP-2013-24}\\
\end{flushright}
\vskip 8pt
{\bf  Self--Dual Conformal Gravity }\vskip 3pt}
\author{Maciej\ Dunajski
\thanks{email {\tt m.dunajski@damtp.cam.ac.uk}}\\
{\sl Department of Applied Mathematics and Theoretical Physics} \\
{\sl University of Cambridge} \\
{\sl Wilberforce Road, Cambridge CB3 0WA, UK.} \\
Paul \ Tod \thanks{email {\tt paul.tod@sjc.ox.ac.uk}}\\
{\sl The Mathematical Institute, Oxford University  }\\
{\sl 24-29 St Giles, Oxford OX1 3LB, UK.}}
\date{1 October 2013}
\maketitle
\begin{abstract}
We find necessary and sufficient conditions for a Riemannian four--dimensional manifold $(M, g)$ with anti-self-dual Weyl tensor to be locally
conformal to a Ricci--flat manifold. These conditions are expressed as the vanishing of scalar and tensor conformal invariants. The invariants  
obstruct the existence
of parallel sections of a certain connection on a complex rank-four vector bundle over $M$. They provide a natural generalisation of the Bach tensor which vanishes identically for anti-self-dual conformal structures. 
We use the obstructions to demonstrate that
LeBrun's anti-self-dual metrics on connected sums of $\CP^2$s are not conformally Ricci-flat on any open set. 

We analyze both Riemannian and neutral signature metrics. In the latter case we find all
anti-self-dual metrics with a parallel real spinor which are locally conformal to Einstein metrics with non--zero cosmological constant. These metrics admit a hyper-surface orthogonal null Killing vector and thus give rise
to projective structures on the space of $\beta$--surfaces. 

\end{abstract}

\section{Introduction}
Let $(M, g)$ be a Riemannian four--manifold, and let $C_{abcd}$ and $R_{ab}$
be respectively the Weyl tensor and the Ricci tensor\footnote{We use the abstract index notation \cite{PR}. Tensors with upper/lower indices
$a, b, \dots=1, \dots, 4 $ are sections of powers of tangent/cotanget
bundle. We employ the summation convention, so $|X|^2=g_{ab}X^aX^b$ denotes the squared length of a vector $X^a$. The isomorphism $TM\otimes\C\cong \spp\otimes
\spp'$ identifies a vector index with a pair of spinor indices
$a=AA'$, where $A, B, \dots =0, 1$ and $A', B', \dots =0, 1$. The anti--self--dual Weyl tensor 
$C_{abcd}$ has a spinor decomposition
$
C_{abcd}=\psi_{ABCD}\epsilon_{A'B'}\epsilon_{C'D'},
$
where $\psi_{ABCD}=\psi_{(ABCD)}$  is a section of $\mbox{Sym}^4(\spp)$, and $\epsilon'$ is a symplectic stucture on $\spp'$.}
of $g$. The field equations of conformal gravity are the vanishing of the Bach tensor
\[
B_{bc}=\Big(\nabla^{a}\nabla^d-\frac{1}{2}R^{ad}\Big)C_{abcd}.
\]
These equations arise from the action given by the squared norm of the Weyl 
tensor. They are invariant under the conformal rescallings of the metric
$g_{ab}\rightarrow \Omega^2g_{ab}$, and are fourth order PDEs in the metric
coefficients.

 The condition $B_{bc}=0$ is necessary for the metric
$g$ to be conformally related to a Ricci-flat metric \cite{KNT,merkulov,BM,NG}. 
This condition is far from sufficient, and in any signature there exist
metrics which are Bach-flat but not conformal to vacuum (see, e. g. \cite{lewandowski} and \cite{pope13}). In \cite{Mald} (see also \cite{AM})
it was demonstrated that imposing a simple Neumann boundary condition selects
Einstein metrics out of other solutions to the Bach-flat condition. One can instead
look for additional local obstructions to the existence of a Ricci-flat metric
in a Bach-flat conformal class $[g]$. If the Weyl curvature of $[g]$ is 
algebraically general, then $[g]$ contains an Einstein metric, possibly with non-zero scalar curvature,
iff the Eastwood--Dighton tensor vanishes, i. e. $E_{abc}=0$, 
where
\[
E_{abc}:=\psi_{ABCD}\nabla^{DD'}\psi_{A'B'C'D'}-
\psi_{A'B'C'D'}\nabla^{DD'}{\psi}_{ABCD},
\]
and $\psi_{ABCD}$ and $\psi_{A'B'C'D'}$ are anti-self-dual
(ASD) and self-dual (SD) Weyl spinors respectively.

If the Weyl tensor of $[g]$ 
is anti--self--dual then both $B_{ab}$ and $E_{abc}$ vanish 
identically\footnote{The vanishing of $E_{abc}$ is obvious,
as the ${\psi}_{A'B'C'D'}=0$. 
The Bach tensor is equal to
\begin{eqnarray*}
B_{ab}&=&2({\nabla^C}_{A'}{\nabla^D}_{B'}+{\Phi^{CD}}_{A'B'})\psi_{ABCD}\\
&=&2({\nabla^{C'}}_{A}{\nabla^{D'}}_{B}+{\Phi^{C'D'}}_{AB})\psi_{A'B'C'D'},
\end{eqnarray*}
where $\Phi_{ABA'B'}$ is the traceless Ricci spinor. Thus 
$B_{ab}=0$ in both SD and ASD cases.}. Therefore any conformally ASD
metric is a solution to the conformal gravity equations. 
In this paper we find necessary and sufficient conditions for a four--dimensional manifold $(M, g)$ with anti-self-dual Weyl tensor to be locally
conformal to a Ricci--flat manifold. These conditions are expressed as the vanishing of one scalar invariant, and one rank-two tensor invariant constructed  from the Schouten tensor,
the conformal curvature and its covariant derivatives. The invariants 
obstruct the existence
of parallel sections of a certain connection on a natural rank-four complex vector bundle 
$E=\spp'\oplus\spp$
over $M$, where $\spp$ and $\spp'$ are complex rank-two symplectic vector bundles (the spin bundles) over $M$ arising from the canonical isomorphism $TM\otimes\C=\spp\otimes\spp'$. Such local obstructions  arise because the conformal to ASD Ricci--flat problem leads
to an over--determined system of PDEs (\ref{twist_eq}) of finite type \cite{eastwood_finite}. 

In Proposition \ref{prop_twist_eq} we shall 
establish the relationship between
existence of a Ricci--flat metric in an ASD conformal class, and existence of a two-dimensional 
vector space of parallel sections of a 
connection ${\mathcal D}$ on  $E$ given by
\be
\label{connection}
{\quad{\mathcal{D}} \left(\begin{array}{c}
\pi\\ 
\alpha
\end{array} \right)= 
\left(\begin{array}{c} \nabla\pi-\alpha\otimes\epsilon \\ 
\nabla\alpha+\pi\hook P
\end{array} \right).}
\ee
Here $\nabla$ is the spin connection on $\spp$ and $\spp'$ induced by the Levi--Civita connection, $P_{ab}=(1/12) R g_{ab}-(1/2)R_{ab} $ is the 
Schouten tensor,  
$\epsilon$ and $\epsilon'$ are symplectic structures on $\spp$ and $\spp'$ respectively
such that $g=\epsilon\otimes\epsilon'$ ,
and $\pi\in\Gamma(\spp'), \alpha\in\Gamma(\spp)$. Finally $\hook$ is denotes contraction
with a section of $\spp'$. The connection ${\mathcal D}$ will arise as the prolongation connection for the twistor equation (\ref{twist_eq}).

We shall analyze both the Riemannian and the neutral signature metrics. 
The emphasis is on the Riemannian case, where the complete characterization of conformal to vacuum condition can be achieved by constructing a rank-two tensor $T_{ab}$ on $M$
which is conformally invariant, i.e. $T\rightarrow T$ when $g\rightarrow \Omega^2 g$,
and vanishes for ASD metrics conformal to vacuum. To formulate our main result 
define
$V\in \Gamma(T^*M)$  by
\be
\label{tensor_V}
V_a=\frac{4}{|C|^2}{C^{bcd}}_a\nabla^eC_{bcde},
\ee
where $|C|^2=C_{abcd}C^{abcd}$.
\begin{theo}
\label{theo_main_th}
Let $g$ be a Riemannian metric with anti-self-dual conformal curvature. Then the conformal class
of $g$ contains a Ricci-flat metric if and only if
\be
\label{main_th_1}
4\nabla^eC_{bcde}\nabla_f C^{bcdf}-|V|^2|C|^2=0
\ee
and 
\be
\label{Tab_tensor}
T_{ab}:=P_{ab}+\nabla_a V_b+V_aV_b-\frac{1}{2}|V|^2 g_{ab}=0.
\ee
\end{theo}
(In applications, and when implementing the calculation of obstructions on a computer, one would clear denominators in these expressions). 
The scalar invariant (\ref{main_th_1}) will be constructed in Section 
\ref{sec_determinants} (Proposition \ref{prop_riem}) as the determinant of a four by four matrix\footnote{Using the spinor notation we would define
$W\in\Gamma(\spp'\otimes\mbox{Sym}^3(\spp))$ and
$V\in \Gamma(T^*M)$  by
\[
W_{A'ABC}={\nabla_{A'}}^D\psi_{ABCD}, \quad
V_{AA'}=\frac{2}{|\psi|^2}\psi_A^{\;\;\;BCD}\nabla_{A'}^E\psi_{BCDE},
\]
where $\psi_{ABCD}$ is the ASD Weyl spinor. The scalar invariant (\ref{main_th_1})
then takes the form
\[
2|W|^2-|\psi|^2|V|^2=0.
\]
We note that the LHS is proportional to  the total square of the Bailey--Eastwood invariant \cite{BE}.
}
whose kernel contains a parallel section of ${\mathcal D}$.
The tensor (\ref{Tab_tensor}) will be constructed from a covariant derivative
of the curvature of ${\mathcal D}$ in Section \ref{main_section}, where we shall  prove Theorem (\ref{theo_main_th}).
We shall also show that ${\mathcal D}$ is a solution to the anti--self--dual Yang--Mills equations on $M$,
and thus, by Ward's twistor transform \cite{ward}, it corresponds to a rank--four holomorphic vector bundle over the twistor space of $(M, g)$.

In Section \ref{sec_examples} we give several examples of ASD conformal structures
without Ricci-flat metrics in the conformal class. For example, we show (Theorem \ref{thmlb}) that LeBrun's metrics on connected sums of several copies of $\CP^2$s are not conformally Ricci-flat on any open set. 
We also show that the Taub-NUT metric with negative mass
is contained in the conformal class of a limiting case of LeBrun's scalar-flat K\"ahler metric
on a line bundle over $\CP^1$ with negative Chern number. 

 In Section \ref{sec22} we shall characterize the anti-self-dual metrics 
in neutral signature which admit  a
parallel real spinor and which are locally conformal to Einstein metrics with non--zero cosmological constant. These metrics admit a hyper-surface orthogonal null Killing vector and thus give rise 
to projective structures \cite{DW07} on the space of $\beta$--surfaces. 

\vspace{2ex}{\bf Acknowledgements.} 
We are grateful to  Gary Gibbons for helpful discussions, and to the anonymous referees for remarks which have resulted in improvements of the manuscript.
\section{The twistor equation}
Let $g$ be a (pseudo) Riemannian metric on an oriented four--dimensional manifold $M$, and let
$[g]=\{\Omega^2 g| \Omega:M\rightarrow \R^+\}$ be the conformal class of metrics containing $g$.
We shall assume that the Weyl tensor of $g$ is anti--self--dual, i. e.
\[
C_{abcd}=-\frac{1}{2}{\epsilon_{ab}}^{ef}C_{cdef}.
\]
The ASD property is conformally invariant, so the Weyl tensor of any metric in $[g]$
is also ASD. If the signature of $g$ is Lorentzian, then (if $[g]$ is ASD) the Weyl tensor is necessarily zero
and the conformal class $[g]$ is flat. Therefore from now on  we shall assume that the signature of $g$ is Riemannian, or neutral. 
\subsubsection*{Riemannian signature} In the former case locally there exist
complex rank two vector bundles $\spp$ and $\spp'$ over $M$ equipped with covariantly
constant symplectic structures $\epsilon$ and $\epsilon'$ 
such that
\be
\label{can_bun_iso}
\C\otimes T M\cong {\spp}\otimes {\spp'}
\ee
is a  canonical bundle isomorphism, and
\[
g(v_1\otimes w_1,v_2\otimes w_2)
=\epsilon(v_1,v_2)\epsilon'(w_1, w_2)
\]
for $v_1, v_2\in \Gamma(\spp)$ and $w_1, w_2\in \Gamma(\spp')$.
We use the conventions \cite{PR} where the spinor indices are capital
letters, unprimed for sections of $\spp$ and primed for sections of $\spp'$. For example $\mu_{A}$ denotes a section of $\spp^{*}$, the dual of $\spp$, and $\nu^{A'}$ a section of $\spp'$.
The symplectic structures $\epsilon_{AB}$
and $\epsilon_{A'B'}$ (such that
$\epsilon_{01}=\epsilon_{0'1'}=1$) are used to lower and
raise the spinor indices according to $\mu_{A}:=\mu^{B}\epsilon_{BA}, 
\mu^A=\epsilon^{AB}\mu_B$. In Riemannian signature, complex conjugation maps $\spp'$ (respectively
$\spp$) to itself by $\pi_{A'}=(p, q)\rightarrow \pi^{\dagger}_{A'}=(-\ov{q}, \ov{p})$ so that the square
of conjugation is minus the identity endomorphism. Thus there is no invariant
notion of real spinors in this case.
\subsubsection*{Neutral signature} If the signature of $(M, g)$ is neutral there exists a notion of real spinors, and the decomposition of the tangent bundle takes the form
\[
TM=\spp\otimes\spp',
\]
where now $\spp$ and $\spp'$ are real rank-two symplectic vector bundles. One can of course also introduce
complex spinors (\ref{can_bun_iso}) and define the fibres of the real spinor bundles
as fixed sets of the conjugation $\pi_{A'}=(p, q)\rightarrow \ov{\pi}_{A'}=(\ov{p}, \ov{q})$. We shall slightly
abuse notation and denote the complex and real spinor bundles by the same symbols.
\subsubsection*{Curvature decomposition and conformal transformations}
The spinor decomposition 
of the Riemann tensor of $g$ is
\begin{eqnarray} \label{riemann}
R_{abcd} &=& \psi_{ABCD} \epsilon_{A'B'} \epsilon_{C'D'} + {\psi}_{A'B'C'D'}
\epsilon_{AB} \epsilon_{CD} \nonumber\\ 
&&+ \Phi_{ABC'D'} \epsilon_{A'B'}\epsilon_{CD} +
\Phi_{A'B'CD} \epsilon_{AB}\epsilon_{C'D'} \nonumber\\&& + 2\Lambda (\epsilon_{AC}\epsilon_{BD}
\epsilon_{A'B'} \epsilon_{C'D'} - \epsilon_{AB} \epsilon_{CD} \epsilon_{A'D'} \epsilon_{B'C'}),
\end{eqnarray}
where $\psi_{ABCD}$ and $\psi_{A'B'C'D'}$ are ASD and SD Weyl spinors
which are symmetric in their indices, 
$\Phi_{A'B'CD}=\Phi_{(A'B')(CD)}$ is the traceless Ricci spinor and $\Lambda=R/24$ is the cosmological constant.
 The spinor Ricci identities 
\be
\label{cc1}
{\nabla^A}_{(A'}\nabla_{B')A}\alpha_B+\Phi_{A'B'AB}\alpha^A=0,
\ee
\[
{\nabla^{A'}}_{(A}\nabla_{B)A'}\beta_{B'}+\Phi_{A'B'AB}\beta^{A'}=0,
\]
\[
{\nabla^{A'}}_{(A}\nabla_{B)A'}\alpha_{C}+\psi_{ABCD}\alpha^D- 2\Lambda\alpha_{(A}\epsilon_{B)C}=0,
\]
\[
{\nabla^{A}}_{(A'}\nabla_{B')A}\beta_{C'}+\psi_{A'B'C'D'}\beta^{D'}- 2\Lambda\beta_{(A'}\epsilon_{B')C'}=0,
\]
hold for any $\alpha\in \Gamma(\spp)$ and $\beta\in \Gamma(\spp')$.

Under conformal rescaling
\[
\hat{g}_{ab}=\Omega^2 g_{ab}
\]
we have
\[
\hat{\epsilon}_{AB}= \Omega\epsilon_{AB}, \quad
\hat{\epsilon}^{AB}= \Omega^{-1}\epsilon^{AB}
\]
and
\[
\hat{\psi}_{ABCD}= \psi_{ABCD}, \quad
\hat{\psi}^{ABCD}= \Omega^{-4}\psi^{ABCD}
\]
with analogous formulae for primed spinors.
Setting $\Upsilon_a=\Omega^{-1}\nabla_a\Omega$ we also find
\begin{eqnarray*}
\hat{\nabla}_{A'B}\hat{\psi}_{CDEA}&=&
\nabla_{A'B}{\psi}_{CDEA}\\
&&-\Upsilon_{A'C}{\psi}_{BDEA}
-\Upsilon_{A'D}{\psi}_{BCEA}
-\Upsilon_{A'E}{\psi}_{BCDA}
-\Upsilon_{A'A}{\psi}_{BCDE}.
\end{eqnarray*}
The Ricci spinor does not have particularly `nice' conformal properties, but its modification (known as the Schouten tensor)
\[P_{ABA'B'}=\Phi_{ABA'B'}-\Lambda\epsilon_{AB}\epsilon_{A'B'}\]
transforms as
\[
\hat{P}_{ab}=
P_{ab}-\nabla_a\Upsilon_b+\Upsilon_a\Upsilon_b-\frac12g_{ab}\Upsilon_c\Upsilon^c.
\]
\subsubsection*{The twistor equation}
Our first result is a characterisation of conformal classes containing Ricci-flat metrics in terms of solutions to 
the valence-one twistor equation. The following Proposition applies to both Riemannian and neutral
signatures.
\begin{prop}
\label{prop_twist_eq}
A metric $g$ with ASD conformal curvature is conformal to a Ricci-flat metric if and only if
there exist two linearly independent solutions to the twistor equation
\be
\label{twist_eq}
\nabla_{A(A'}\pi_{B')}=0.
\ee
\end{prop}
{\bf Proof.}
First assume that $g$ is conformally equivalent to a Ricci-flat metric $\hat{g}=\Omega^2 g$, where
$\Omega$ is a non--zero function on $M$. Therefore the Riemann tensor of 
$\hat{g}$ is anti--self--dual and thus the spinor connection on $\spp'$ is flat. Therefore there exists
a basis of covariantly constant spinors, say $\hat{\pi}_{A'}$ and $\hat{\mu}_{A'}$, on $\spp'$. The twistor equation
is conformally invariant: if $\hat{\pi}_{A'}=\Omega\pi_{A'}$  
then $\hat{\nabla}_{A(A'} \hat{\pi}_{B')}=\Omega \nabla_{A(A'}\pi_{B')}$. Thus $(\pi_{A'}, \mu_{A'})$ is a pair of linearly
independent solutions to (\ref{twist_eq}).

Conversely, assume that $g$ admits two linearly independent
solutions to (\ref{twist_eq}).
The condition (\ref{twist_eq}) is equivalent to 
\be\label{t1} \nabla_{AA'}\pi_{B'}=\epsilon_{A'B'}\alpha_A\ee 
for some section $\alpha_A$ of $\spp$.
Given a
solution, recall the conformal transformations:
\be
\label{conf_change_pi}
\hat{g}_{ab}=\Omega^2g_{ab},\;\;\hat{\pi}_{A'}=\Omega\pi_{A'},\;\;\hat{\alpha}_A=\alpha_A+\Upsilon_{AC'}\pi^{C'},
\ee
where $\Upsilon_a=\Omega^{-1}\nabla_a\Omega$.
Thus if $\psi_{A'B'C'D'}=0$ and $g$ is 
conformal to vacuum, there will be a two--dimensional vector space of
solutions to (\ref{t1}). This will turn out to be sufficient as
well. To see it commute derivatives on (\ref{t1}) to deduce \be\label{t2}
\nabla_{AA'}\alpha_B=-P_{ABA'B'}\pi^{B'},\ee where
\[P_{ABA'B'}=\Phi_{ABA'B'}-\Lambda\epsilon_{AB}\epsilon_{A'B'}\]
is the Schouten tensor.
Now commute derivatives on (\ref{t2}). From ${\nabla^{B}}_{(A'}\nabla_{B')B}\alpha_A$
we get an identity after using the Bianchi identities
\[\nabla^A_{(A'}\Phi_{B')C'AB}+\nabla_{B(A'}\Lambda\epsilon_{B')C'}=0,\]
which in turn uses the vanishing of the SD Weyl spinor. However from 
${\nabla^{A'}}_{(A}\nabla_{B)A'}\alpha_C$ we
obtain a condition:
\be\label{c1}\psi_{ABCD}\alpha^D-\nabla^{A'}_{(A}\Phi_{BC)A'B'}\pi^{B'}=0.\ee
This can be rewritten using the Bianchi identity as
\be\label{c2}\psi_{ABCD}\alpha^D-\nabla^{F}_{A'}\psi_{ABCF}\pi^{A'}=0.\ee
This is a set of four linear equations in four unknowns and so has a
solution provided the determinant of the associated four by four matrix is zero. We shall investigate this determinant condition
in the next Section.
To justify the claim of sufficiency, suppose $(\pi_{A'},\alpha_A)$
and $(\mu_{A'},\beta_A)$ are two linearly independent solutions of (\ref{t1}) and 
(\ref{t2}). Provided the inner
product is nonzero (which readily follows from linear independence)
introduce
\be
\label{conf_factor}
\Omega=(\pi_{A'}\mu^{A'})^{-1}.
\ee
Then
\[-\Omega^{-2}\nabla_a\Omega=\pi_{A'}\beta_A-\mu_{A'}\alpha_A=-\Omega^{-1}\Upsilon_a,\]
and differentiate again
\[\nabla_a(-\Omega^{-1}\Upsilon_b)=\nabla_a(\pi_{B'}\beta_B-\mu_{B'}\alpha_B)\]
\[=g_{ab}\alpha_C\beta^C-\Omega^{-1}P_{ab}.\]
Now notice that
\[\Upsilon_c\Upsilon^c=2\Omega\alpha_C\beta^C,\]
to conclude that
\[P_{ab}-\nabla_a\Upsilon_b+\Upsilon_a\Upsilon_b-\frac12g_{ab}\Upsilon_c\Upsilon^c=0,\]
i.e. $g$ is conformal to vacuum.
\koniec

In the real Riemannian case it is enough to have one solution of
(\ref{t1}) as the Hermitian conjugate then gives another. In the neutral case, where the signature of $g$ is $(2, 2)$ 
there is an invariant notion of real spinors and (\ref{twist_eq}) may have only one solution. We shall analyze this situation in Section \ref{sec22}.

\section{Minors, determinants and the necessary conditions}
\label{sec_determinants}
Set $W_{A'ABC}={\nabla_{A'}}^D\psi_{ABCD}$ and consider the 4 by 4 matrix
\be
\label{sys2}
{\mathcal R} =\left( \begin{array}{cccc}
 \psi_{0000}\;\ \psi_{0001} & -W_{0'000} & -W_{1'000} \\
 \psi_{0001}\;\ \psi_{0011} & -W_{0'001} & -W_{1'001} \\
 \psi_{0011}\;\ \psi_{0111} & -W_{0'011} & -W_{1'011} \\
 \psi_{0111}\;\ \psi_{1111} & -W_{0'111} & -W_{1'111} 
 \end{array} \right). 
\ee
Equation (\ref{c2}) can be written as ${\mathcal R} \;\Psi=0$, where $\Psi= (\alpha^0, \alpha^1, \pi^{0'}, \pi^{1'})^T$=0.
We find 
\begin{eqnarray}
\label{det_M}
\mbox{det}\;({\mathcal R})&=&|\psi|^2|W|^2-2\psi^{EFGH}\psi_{ABCH}W^{A'ABC}W_{A'EFG}\\\nonumber
&=&\frac{1}{2}|\psi|^2(2|W|^2-|\psi|^2|V|^2)=0,
\end{eqnarray}
where $V_a=2|\psi|^{-2}{\psi_A}^{BCD}W_{A'BCD}$, or equivalently in the ASD case, $V_a$ is given by
(\ref{tensor_V}).
Proposition \ref{prop_twist_eq} implies that
the determinant (\ref{det_M}) must vanish for ASD conformal structures containing a Ricci-flat metric. To have two linearly independent solutions
the matrix ${\mathcal R}$ must have rank at most two (we shall deal with the rank one case in Proposition \ref{type_N}). For that we need 
all sixteen 3 by 3 minors of ${\mathcal R}$ to vanish. Eight of these are
\be
\label{BE_inv}
K_{A'ABC}:=|\psi|^2W_{A'ABC}-2\psi^{EFGH}\psi_{ABCH} W_{A'EFG}=0.
\ee
These are the Bailey--Eastwood invariants \cite{BE}. They can be derived directly from
(\ref{c2}): solve it for $\alpha^A$ to find $\alpha^{A}={V^A}_{A'}\pi^{A'}$, and substitute it back
to (\ref{c2}). This gives a linear constraint $K_{A'ABC}\pi^{A'}=0$ on the initial data
for the twistor equation. The data should be specifiable freely, so (\ref{BE_inv}) follows.

The eight remaining minors are
\be
\label{L_minors}
L_{ABCD}:=|W|^2\psi_{ABCD}-2\psi_{EFGD}W^{A'EFG}W_{A'ABC}=0.
\ee
Note that the RHS is a section of $(\spp\odot\spp\odot\spp)\otimes\spp$. Using
\[
\mbox{Sym}^3(\C^2)\otimes\C^2=\mbox{Sym}^4(\C^2)\oplus \mbox{Sym}^4(\C^2),
\]
and $\mbox{Sym}^k(\C^2)=\C^{k+1}$ we can decompose the invariants (\ref{L_minors}) 
in the representation theoretic way into 5+3 irreducible conditions
\[
S_{ABCD}:=L_{(ABCD)}=0,\quad N_{AB}:={\psi_{EFG}}^DW^{A'EFG}W_{A'ABD}=0.
\]
So we have proved 
\begin{prop}
\label{prop_KSN}
Necessary conditions for an ASD conformal class
to contain a Ricci-flat metric are (\ref{BE_inv}) and (\ref{L_minors}) or
\[
K=S=N=0.
\]
All these conditions are third order in the metric.
\end{prop}
The spinor conditions from Proposition (\ref{prop_KSN}) correspond to tensors:
\[
N_{ab}=N_{AB}\epsilon_{A'B'}, \quad K_{abc}=-4 K_{A'ABC}\epsilon_{B'C'}
\] and 
$S_{abcd}=S_{ABCD}\epsilon_{A'B'}\epsilon_{C'D'}$ where
\[
K_{abc}=|C|^2\nabla^d C_{abcd}-4C^{efgh}C_{abch}\nabla^d C_{efgd}, \quad
N_{ab}=C^{eqcd}\nabla^pC_{pqcd}\nabla^fC_{feab}.
\]
Finally we verify by explicit calculation than the scalar invariant (\ref{det_M})
is proportional to the spinor (or tensor) norms of both $K_{A'ABC}$ and $L_{ABCD}$:
\[
|K|^2= |\psi|^2\mbox{det}\;({\mathcal R}), \quad |L|^2=|W|^2 \mbox{det}\;({\mathcal R}).
\] 
\subsection{Riemannian signature}
We shall now consider Riemannian signature and 
show that in this case all 3 by 3 minors from Proposition (\ref{prop_KSN})
vanish identically if $\mbox{det}({\mathcal R})=0$. Equivalently, we shall show that if rank ${\mathcal R}$ is smaller than four, then it is at most two. It will then follow from Proposition
\ref{type_N} that if $\mbox{det}{({\mathcal R})}=0$, the signature of $g$ is Riemannian, and $g$ is not conformally flat,
then rank ${\mathcal R}$ is exactly two.
\begin{prop}
\label{prop_riem}
In the Riemannian case the sixteen conditions
(\ref{BE_inv}) and (\ref{L_minors}) hold identically if $\mbox{det}({\mathcal R})=0$, i.e. if {\em(}\ref{det_M}{\em)} holds.
\end{prop}
{\bf Proof.}
Set
\be
\label{NP}
\psi_0=\psi_{0000},\quad \psi_1=\psi_{0001},\quad \psi_2=\psi_{0011},\quad \psi_3=\psi_{0111},\quad \psi_4=\psi_{1111}.
\ee
We can chose a basis of $\spp$ which consists of a spinor together with its Hermitian conjugate, and do the same for $\spp'$. 
The Riemannian reality conditions
on spinors then yield
\[
{\psi_3}=-\ov{\psi_1}, \quad \psi_4=\ov{\psi_0}, \quad \ov{\psi_2}=\psi_2
\]
and
\[
W_{0'000}=\ov{W_{1'111}}, \quad W_{0'001}=-\ov{W_{1'110}},\quad W_{0'011}=\ov{W_{1'100}},\quad W_{0'111}=-\ov{W_{1'000}}. 
\]
Set 
\[
w_0=-W_{0'000}, \quad w_1=-W_{0'001}, \quad w_2=-W_{0'011}, \quad w_3=-W_{0'111}.
\]
It is always possible to perform an $SU(2)$ rotation on $\spp$  to chose the unprimed spin frame (a basis of $\spp$) such that $\psi_0=0$, and perform
an independent $SU(2)$ rotation of $\spp'$ to  chose
the primed spin frame (a basis of $\spp'$) such that $W_{0'000}=0$. Then
\be
\label{sys2_new}
{\mathcal R}=
\left(
\begin{array}{cccc}
 0 & \psi_1 & 0 & -\ov{w_3}\\
 \psi_1 & \psi_2 & w_1 & \ov{w_2} \\
 \psi_2 & -\ov{\psi_1} & w_2 & -\ov{w_1}\\
 -\ov{\psi_1} & 0 & w_3 & 0 
\end{array}
\right) 
\ee
and the determinant is a sum of two non-negative numbers
\[
\mbox{det}({\mathcal R})=|\ov{\psi_1}\;w_1+\psi_1 \;w_3|^2+|\psi_2 \;w_3+\ov{\psi_1}\;w_2|^2.
\]
This vanishes if and only if four real quadratic conditions (which we write as two complex equations) hold
\be
\label{det_minors}
\ov{\psi_1}\;w_1+\psi_1 \;w_3=0, \quad \psi_2\; w_3+\ov{\psi_1}\;w_2=0.
\ee
We verify by explicit calculation
that vanishing of these four quadratics is equivalent to the vanishing of the sixteen cubics
which give the minor conditions (\ref{BE_inv}) and (\ref{L_minors}). 
\koniec
\subsection{Neutral signature}
Vanishing of (\ref{BE_inv}) and (\ref{L_minors})
guarantees that the rank of the matrix ${\mathcal R}$ is at most two. 
Now we shall consider the rank-one case, and show that in this case the signature of $g$ is necessarily neutral,
and the conformal curvature is of type $N$, i.e. the homogeneous quartic
$\psi_{ABCD}z^Az^Bz^Cz^D$ has a root of multiplicity four \cite{PR}. 
\begin{prop}
\label{type_N}
Let ${\mathcal R}$ be the four by four matrix (\ref{sys2}). If rank $({\mathcal R})=1$ then the signature of $g$ is $(2, 2)$ and the ASD 
Weyl tensor is of type $N$.
\end{prop}
{\bf Proof.}
Assume that $\mbox{rank}\;({\mathcal R})=1$. In this case the first two columns of ${\mathcal R}$ have to be linearly dependent.
Therefore
\[
\psi_1=\lambda \psi_0, \quad\psi_2=\lambda\psi_1,\quad \psi_3=\lambda\psi_2,\quad \psi_4=\lambda\psi_3,
\]
and, using $o_A, \iota_{A}$ as a basis of $\Lambda(\spp^*)$,
\[
\psi_{ABCD}=(\iota_{A}\iota_{B}\iota_C\iota_D-4\lambda \iota_{(A}\iota_B\iota_C o_{D)}
+6\lambda^2\iota_{(A}\iota_B o_C o_{D)}- 4\lambda^3 \iota_{(A} o_B o_C o_{D)}
+\lambda^4 o_Ao_Bo_C o_D)\psi_0.
\]
Now change the spin frame by $\hat{\iota}_A=\iota_A-\lambda o_A, \hat{o}_A=o_A$
so that $\psi_{ABCD}=\hat{\iota}_A \hat{\iota}_B\hat{\iota}_C\hat{\iota}_D\;\psi_0$,
and the ASD Weyl spinor is of  Petrov--Penrose type $N$. This can only happen
in signature (2, 2). In Riemannian signature the roots of the Weyl quartics are distinct,
or come in two pairs of repeated roots (type $D$).
\koniec
This result combined with Proposition \ref{prop_riem} gives the following
\begin{col}
\label{Coroll}
Let $g$ be a Riemannian metric with ASD conformal curvature which is not conformally flat, and such that $\mbox{det}\; ({\mathcal R})=0$, where
${\mathcal R}$ is given by (\ref{sys2}). Then rank$({\mathcal R})=2$.
\end{col}
\section{Twistor curvature and sufficient conditions}
\label{main_section}
The conditions $K=0$ and $L=0$, given by (\ref{BE_inv}) and 
(\ref{L_minors}) are clearly necessary for the existence
of a Ricci-flat metric in the ASD conformal class. They are however not 
sufficient, and in this Section we shall establish sufficient 
conditions in the Riemannian signature. Our method follows the approach of
\cite{BDE} to the metrisability problem in projective geometry.
Let us recall that the necessary conditions arise from imposing a rank-2 condition on a four by four matrix ${\mathcal R}$ given by (\ref{sys2}). If the rank of this matrix is two (which  in the Riemannian signature is guaranteed by the 
vanishing of the determinant (\ref{det_M}) - see Corollary \ref{Coroll})
then the linear system of equations (\ref{c2}) admits two linearly 
independent solutions. We do however need to make sure that these solutions 
satisfy the linear system of PDEs (\ref{t1}) and (\ref{t2}). To do this, we 
differentiate (\ref{c2}) covariantly, and eliminate the derivatives 
$\nabla\pi$, and $\nabla\alpha$ using (\ref{t1}) and (\ref{t2}). 
This leads to more linear homogeneous equations on $(\alpha_0, \alpha_1, \pi_{0'},
\pi_{1'})$. These equations must be satisfied identically as a consequence
of   (\ref{BE_inv}) and (\ref{L_minors}) (or, in Riemannian signature
as a consequence of (\ref{det_M})) as otherwise there would be more than 
two independent equations on four unknowns, and  two independent solutions
to (\ref{t1}) and (\ref{t2}) would not exist. This,  
by Proposition \ref{prop_twist_eq}, would imply that the ASD conformal 
structure does not contain a Ricci-flat metric.

In the construction below we shall first 
reformulate (\ref{t1}) and (\ref{t2}) in terms of a  connection on a rank four vector bundle over $M$, and then implement the differentiation procedure/adding
new homogeneous equations in terms of restrictions on the holonomy of 
this connection. Moreover we shall restrict ourselves to Riemannian 
signature, as there the analysis of the sufficient conditions is particularly simple: if the necessary conditions hold then the 
rank of the matrix (\ref{sys2}) is two, so there are exactly two independent 
conditions on $\pi_{A'}$ and $\alpha_{A}$. If one differentiation
of  (\ref{c2}) does not give new conditions (which will be the case
if the obstructions from Theorem \ref{main_thm}) vanish), then the rank of the 
matrix constructed by all homogeneous linear constraints is still two.
This means that no new conditions would be added by subsequent 
differentiations, and we can stop the process of adjoining equations. 
The space of solutions to  (\ref{t1}) and (\ref{t2}) is then,
by the Frobenius theorem, two--dimensional\footnote{The analysis would not be
so simple if we allowed neutral signature, as then the rank of
the matrix (\ref{sys2}) could be one if $[g]$ were of type $N$ as in 
Proposition \ref{type_N}. The rank could then go up to two after one 
differentiation, and we would need to differentiate once more to ensure
that the rank does not increase. Thus, in the neutral signature
case the sufficient conditions are given by expressions which are third order
in the conformal curvature, and so fifth order in the metric.}. 

Using the method outlined above we shall establish
\begin{theo}
\label{main_thm}
Let $g$ be a Riemannian metric with ASD conformal curvature. Then the conformal class
of $g$ contains a Ricci-flat metric if and only if 
$\mbox{det}({\mathcal R})$ given by
(\ref{det_M}) vanishes, and
\be
\label{new_invariant}
\Xi_{E'EABCD}{V^{D}}_{A'}-\Theta_{ABCEE'A'}=0,  
\ee
where $W_{A'ABC}={\nabla_{A'}}^D\psi_{ABCD},  V_{AA'} =2|\psi|^{-2}{\psi_A}^{BCD}W_{A'BCD}$,
and $(\Xi, \Theta)$ are given by (\ref{xi_form}, \ref{theta_form}) respectively.
\end{theo}
Before giving a proof of this result, we shall develop some formalism.
Consider the rank-four complex vector bundle over $E\rightarrow M$
\be
\label{Ebundle}
E=\spp'\oplus \spp
\ee
with sections $\Psi_{\alpha}$ which under conformal rescalings
of the metric transform like $\Psi_\alpha\rightarrow\hat{\Psi}_\alpha$, where
\be
\label{4-tractor}
\hat{\Psi}_{\alpha}:=
\left(\begin{array}{c}\hat{\pi}_{A'}\\\hat{\alpha}_A
\end{array} \right) 
=
\left(\begin{array}{c}\Omega\pi_{A'}\\ \alpha_{A}+\Upsilon_{AB'}\pi^{B'}
\end{array} \right). 
\ee
Our approach is very much in the spirit of \cite{BEG}. We shall call $E$ the dual twistor bundle, and refer to its sections as dual twistors.
Define a derivative $\mathcal{D}$ on this vector bundle by
\be
\label{tractorD}
{\mathcal{D}}_a{\Psi}_{\beta}:=
\left(\begin{array}{c}
\nabla_{AA'}\pi_{B'}-\epsilon_{A'B'}\alpha_{A}\\
\nabla_{AA'}\alpha_{B}+P_{ABA'B'}\pi^{B'}
\end{array} \right). 
\ee
This connection, together with
the conformal transformation properties (\ref{4-tractor}),
is recognisable as local twistor transport for a dual twistor
(given for twistors in \cite{PR}, p 113).
\begin{prop}
\label{prop_asd_1}

\begin{enumerate}
\item
There is a correspondence between non--zero parallel sections
of the connection (\ref{tractorD}) on the bundle $E$ over a Riemannian
four manifold $(M, g)$ and Ricci-flat metrics in an ASD conformal class $[g]$; if the section is multiplied by a complex constant $\alpha$, then the metric is multiplied by $|\alpha|^{-2}$. Conversely given a Ricci-flat metric in the ASD conformal class, there is a two-dimensional
vector space of parallel sections spanned by one solution and its Hermitian conjugate.
\item The curvature of (\ref{tractorD}) is an ASD Yang--Mills field on $(E, M)$.
\end{enumerate}
\end{prop}
{\bf Proof.}
The first part of the Proposition is an immediate consequence of 
Proposition \ref{prop_twist_eq} and the form of the system (\ref{t1}) and (\ref{t2}).

We now calculate the curvature of the  connection (\ref{tractorD}) from 
the general formula
\[{\mathcal{D}}_{[c}{\mathcal{D}}_{a]}{\Psi}_{\beta}=\frac12{{\mathcal{R}}_{ca\,{\beta}}}^{{\delta}}\Psi_{\delta}
=
\left(\begin{array}{c}
0\\
\psi_{CABD}\alpha^D\epsilon_{C'A'}-W_{D'ABC}\pi^{D'}\epsilon_{C'A'}
\end{array}\right)
\]
and find that it is anti--self--dual on the first pair of indices.
This should be contrasted with the result of Merkulov \cite{merkulov}, where the Bach equations are shown
to arise as full Yang--Mills equations on a connection on $E$.

More explicitly we obtain\footnote{The dual twistor bundle 
indices $\alpha, \beta, \dots=1, \dots, 4$ are identified with the spinor indices
via the isomorphism (\ref{Ebundle}). Using the abstract index notation we write
$\alpha=A+A'$. The vector indices $a, b, \dots =1, \dots, 4$ are identified with the spinor indices by (\ref{can_bun_iso}), i. e. $a=AA'$.}
\[
{{\mathcal{R}}_{ca\beta}}^{\delta}=\epsilon_{C'A'}{{\mathcal{R}}_{CA\beta}}^{\delta}
+\epsilon_{CA}{{\mathcal{R}}_{C'A'\beta}}^{\delta}
\]
where ${{\mathcal{R}}_{C'A'\beta}}^{\delta}=0$,
and
\[
{{\mathcal{R}}_{CAB'}}^{D'}=0,\quad {{\mathcal{R}}_{CAB}}^{D'}={{{W}}_{CAB}}^{D'},\quad
{{\mathcal{R}}_{CAB'}}^{D}=0,\quad 
{{\mathcal{R}}_{CAB}}^{D}={{{\psi}}_{CAB}}^{D}.
\]
\koniec
We have therefore deduced that
$\psi_{ABCD}$ and $W_{A'ABC}$ are components of the dual twistor curvature.
We can thus  identify the 4 by 4 matrix (\ref{sys2})
in the system (\ref{c2}) with the dual twistor curvature.
We shall now prove Theorem \ref{main_thm}, and
find conditions which guarantee that the 
leading spinor part $\pi_{A'}$ of a
dual twistor
$(\pi_{A'}, \alpha_A)$ can be chosen arbitrarily at a point $p$ in $M$
and propagated parallelly with no obstructions to all points in a 
neighbourhood of $p$. This will guarantee the existence of a two-dimensional solution space to the twistor equation (\ref{twist_eq}).\\

{\bf Proof of Theorem \ref{main_thm}}.
We differentiate the four equations (\ref{c2})
and use the parallel condition  on sections 
${\mathcal D}\Psi=0$ to produce a sequence of algebraic matrix equations
\[
{\mathcal R}\Psi=0, \quad  ({\mathcal D}{\mathcal R})\Psi=0, \quad
({\mathcal D}^2{\mathcal R})\Psi=0, \dots\;\;,
\]
where ${\mathcal R}$ is the four by four matrix (\ref{sys2}).
We stop the process once the differentiation does not produce new equations.
This, in the Riemannian signature, happens after just one differentiation.
The rank of ${\mathcal R}$ is two if the necessary conditions
$K=0$ and $L=0$ from Proposition \ref{prop_riem} hold. 
This, in Riemannnian signature, is equivalent to the condition 
(\ref{main_th_1}) - we have shown this in Proposition \ref{prop_riem}.
The rank should not go 
up, or the dimension of the solution space is smaller than two. Therefore
all 3 by 3 minors of the 4 by 20 matrix  $({\mathcal R}, {\mathcal D}{\mathcal R})$ have to vanish. If they do, then we do not need to differentiate 
any more, as subsequent derivatives would only reproduce the homogeneous equations 
${\mathcal R}\Psi=0$. Thus vanishing of the minors will give sufficient conditions for the existence
of a parallel section of ${\mathcal D}$. 

To construct the minors explicitly differentiate ({\ref{c2}}) and eliminate the derivatives using (\ref{t1}) and (\ref{t2}). This will give
a system of equations ${\mathcal D}{\mathcal R}=0$. The result is
\be
\label{2nd_der}
\Xi_{E'EABCD}\alpha^D-\Theta_{ABCEE'A'}\pi^{A'}=0,
\ee
where
\be
\label{xi_form}
\Xi_{E'EABCD}=\nabla_{E'E}\psi_{ABCD}-W_{E'ABC}\;\epsilon_{DE},
\ee
and
\be
\label{theta_form}
\Theta_{ABCEE'A'}=\psi_{ABCD}{P^{D}}_{EE'A'}+\nabla_{EE'}W_{A'ABC}.
\ee

Consider the combined system of linear homogeneous equations (\ref{c2}) and (\ref{2nd_der}).
Only two equations in  (\ref{c2}) are independent as we are assuming that the necessary conditions from Proposition \ref{prop_KSN} hold. We solve these two equations for
$\alpha_{A}=V_{AA'}\pi^{A'}$, where $V_{AA'}$ is given by (\ref{v1}), and substitute $\alpha_A$ into the remaining equations (\ref{2nd_der}) which we insist hold identically.
This gives vanishing of the obstruction (\ref{new_invariant}).
\koniec
\subsection{Main theorem}

The sufficient conditions
(\ref{new_invariant}) are expressed as the vanishing of a section of
\[
\mbox{Sym}^3(\spp)\otimes \spp\otimes \spp'\otimes\spp'.
\]
A general section of this bundle has $32$ independent components, and it is not clear
from the proof of Theorem \ref{main_thm} which of these components are independent.
In this Section we shall establish our main result and show that vanishing of the $32$ conditions (\ref{new_invariant}) is a consequence of vanishing of one rank-two tensor on $M$.

 Multiplying
(\ref{c2}) by $\psi_{ABCE}$ and using $\psi_{ABCD}\psi^{ABCE}=(1/2)|\psi|^2{\delta_D}^E$
we find
\[\alpha_A=V_{AA'}\pi^{A'},\]
with
\be\label{v1}V_{AA'}=\frac{2}{I}\psi_A^{\;\;\;BCD}\nabla_{A'}^E\psi_{BCDE},\ee
and $I=\psi^{ABCD}\psi_{ABCD}$ assuming $I\neq 0$ (which is
always the case in  Riemannian signature\footnote{If signature of $g$ is neutral, and $I=0$ then we solve for $\alpha^A$ using the cubic invariant $J$. If
that is also zero then the conformal curvature is of Petrov--Penrose 
type $N$ or type $III$.}). Note that (\ref{v1}) is the spinor form of
(\ref{tensor_V}).
Now (\ref{t1}) becomes
constancy in a modified connection:
\[D_{AA'}\pi_{B'}:=\nabla_{AA'}\pi_{B'}-\epsilon_{A'B'}V_{AC'}\pi^{C'},\]
with $V_{AA'}$ as in (\ref{v1}). We extend this connection to
a connection on unprimed spinors so that the resulting connection on vectors is torsion-free.
For $g$ to be conformal to vacuum this 
connection must be flat, as only then the initial data $\pi_{A'}|_p$ at a point can be parallelly propagated. This 
leads to a set of conformally invariant obstructions summarised in 
Theorem \ref{theo_main_th}.

{\bf Proof of Theorem \ref{theo_main_th}.}
The necessity of (\ref{main_th_1}) was shown in Proposition \ref{prop_KSN}
and Proposition \ref{prop_riem}. Note that (\ref{main_th_1})
is the tensor form of (\ref{det_M}).

Contracting $\nabla_{AA'} \pi_{B'}=\epsilon_{A'B'} V_{AC'}\pi^{C'}$ with 
${\nabla^{A'}}_{B}$ and using the spinor Ricci identities yields
\[
(P_{ABA'B'}+\nabla_{AA'}V_{BB'}+V_{BA'}V_{AB'})\pi^{B'}=0.
\]
This puts constraints on the initial data for $D_{AA'}\pi_{B'}=0$ unless the expression in bracket vanishes.
This is true, as long as $\pi_{B'}\neq 0$, so we require that (\ref{det_M}) (or equivalently (\ref{main_th_1})) holds.

Rearranging the unprimed indices in $V_{BA'}V_{AB'}$ gives 
\be
\label{t=0}
T_{ab}=0,
\ee
where $T_{ab}$ is given by (\ref{Tab_tensor}),
and in particular
\[
\nabla_{[a}V_{b]}=0, \quad 4\Lambda=\nabla^a V_a- V^a V_a.
\]
To prove the conformal invariance of (\ref{t=0}) use the rules for conformal transformation of the covariant derivative of spinors with $\hat{\psi}_{ABCD}=\psi_{ABCD}$
\[
\hat{\nabla}_{AA'}\hat{\psi}_{BCDE}=\nabla_{AA'}\hat{\psi}_{ABCD}-4\nabla_{A'(E}\psi_{BCD)A}
\]
to show that 
\be
\label{conf_w}
\hat{W}_{A'ABC}=\Omega^{-1}(W_{A'ABC}-\Upsilon_{A'E}{\psi^{E}}_{ABC}) \quad\mbox{and consequently}
\quad \hat{V}_a=V_a+\Upsilon_{a}.
\ee
Now use the conformal transformation of the covariant derivative of one--forms
\[
\hat{\nabla}_a \hat{V}_b=\nabla_a \hat{V}_b-\Upsilon_a \hat{V}_b-\Upsilon_b \hat{V}_a+ g_{ab}\Upsilon^c \hat{V}_c
\]
to find
\[
\hat{\nabla}_a\hat{V}_b+\hat{V}_a\hat{V}_b-\frac{1}{2}\hat{V}_c\hat{V}^c \hat{g}_{ab}
={\nabla}_a{V}_b+{V}_a{V}_b-\frac{1}{2}{V}_c{V}^c {g}_{ab}+
\nabla_a\Upsilon_b-\Upsilon_a\Upsilon_b+\frac12g_{ab}\Upsilon_c\Upsilon^c.
\]
The result now follows by applying the formula for conformal transformation of the Schouten tensor
\[
\hat{P}_{ab}=
P_{ab}-\nabla_a\Upsilon_b+\Upsilon_a\Upsilon_b-\frac12g_{ab}\Upsilon_c\Upsilon^c.\]
We also note that no further conditions arise from contracting $\nabla_{AA'} \pi_{B'}=\epsilon_{A'B'} V_{AC'}\pi^{C'}$ with 
${\nabla^{A}}_{C'}$.

Conversely, if (\ref{t=0}) holds then $\nabla_{[a}V_{b]}=0$ and $V_a$ is locally a gradient.
Thus, by (\ref{conf_w}) it can be set to zero by a conformal rescaling of the metric. 
Now $\hat{\alpha}_A$ in (\ref{conf_change_pi}) vanishes and so Proposition \ref{prop_twist_eq}
implies that $g_{ab}$ is conformal to an ASD Ricci-flat metric.
\koniec
Finally we demonstrate that vanishing of the tensor (\ref{Tab_tensor}) arises directly from
condition (\ref{new_invariant}) in Theorem \ref{main_thm}.
Contracting (\ref{new_invariant}) with ${\psi^{ABC}}_F$  and integrating by parts yields
\[
\Big(\frac{1}{2}\nabla_{EE'}(|\psi|^2\epsilon_{FD})-\psi_{ABCD}\nabla_{EE'}
{\psi^{ABC}}_F\Big)
{V^{D}}_{A'}-\frac{1}{2}|\psi|^2V_{FE'}V_{EA'}
\]
\[
-\frac{1}{2}|\psi|^2P_{FEE'A'}
-\frac{1}{2}\nabla_{EE'}(V_{FA'}|\psi|^2)+W_{A'ABC}\nabla_{EE'}{\psi^{ABC}}_F)=0.
\]
Dividing this expression by $|\psi|^2/2$ and rearranging the remaining terms
gives
\[
P_{FEE'A'}+\nabla_{EE'}V_{FA'}+V_{FE'}V_{EA'}-\frac{2}{|\psi|^4}
K_{A'ABC}\nabla_{EE'}{\psi^{ABC}}_F=0,
\]
where $K_{A'ABC}$ is the Bailey--Eastwood invariant (\ref{BE_inv})
which vanishes as a consequence of (\ref{main_th_1}) and Proposition
\ref{prop_riem}. Rearranging the unprimed indices on $V_{FE'}V_{EA'}$
gives $T_{ab}=0$, where $T_{ab}$ is given by (\ref{Tab_tensor}).



\section{Examples}
\label{sec_examples}
\subsubsection*{LeBrun's anti-self-dual orbifolds}
Consider a two--parameter family of Riemannian metrics
\be
\label{lebrun88}
g=f^{-1} dr^2+\frac{1}{4}r^2(\sigma_1^2+\sigma_2^2 +f \sigma_3^2), \quad\mbox{where}\quad
f=1+\frac{A}{r^2}+\frac{B}{r^4},
\ee
$A, B$ are real constants,
and $\sigma_1, \sigma_2, \sigma_3$ are the left invariant one-forms on the group manifold
$SU(2)$ such that
\[
d\sigma_1=\sigma_2\wedge \sigma_3, \quad  
d\sigma_2=\sigma_3\wedge \sigma_1, \quad
d\sigma_3=\sigma_1\wedge \sigma_2.
\]
These metrics are K\"ahler, with vanishing scalar curvature (therefore they are ASD \cite{Der83})
for any constants $(A, B)$. They arise from the spherically symmetric ansatz on a K\"ahler
potential on an open ball in $\C^2$. If the ratio of the roots of the quadratic $x^2+Ax+B$
is a negative integer $k$, then $g$ is a metric on a holomorphic line bundle 
${\mathcal{O}}(k-1)\rightarrow \CP^1$ with negative Chern class. One-point compactifications
of these asymptotically locally flat manifolds are compact ASD orbifolds which arise as limits of LeBrun's ASD metrics on connected sums of $\CP^2$s, where all points
in LeBrun's hyperbolic ansatz coincide \cite{L91}.

 To look for Ricci-flat metrics in
the conformal class of $g$ we shall examine the invariants of Theorem \ref{theo_main_th} for any value of the constants $A, B$. We find that the one-form $V$ is exact and given by
\[
V=d (\mbox{ln}(Ar^2+2B)).
\]
The scalar invariant (\ref{main_th_1}) vanishes, but the second order obstruction
(\ref{Tab_tensor}) is not identically zero, and is proportional to the metric:
\be
\label{obstruction_T}
T=\frac{A(4B-A^2)}{(Ar^2+2B)^2}\;g.
\ee
This obstruction vanishes when $A=0$. In this case $V=0$, and $g$ is the Ricci-flat
Eguchi--Hanson ALE gravitational instanton. This was already noted in \cite{L88}. However
$T$ also vanishes when 
\[
B=A^2/4.\]
 In this case the quartic $f$ has a repeated root.
The form of $V$ and formula (\ref{conf_w}) gives a conformal factor which makes 
$\hat{g}=\Omega^2\; g$ Ricci-flat, and therefore hyper-K\"ahler. We find
\be
\label{taub_nut}
\hat{g}=\frac{1}{(2r^2+A)^2}\;g.
\ee
Thus both $g$ and $\hat{g}$ are
scalar-flat and K\"ahler, and yet they are conformally related with non-constant conformal factor. The complex structure of $g$ does not belong to the two-sphere
of complex structures of $\hat{g}$. A coordinate transformation
\[
R=\frac{1}{8r^2+4A}\leq \frac{1}{4A}
\]
yields
\[
\hat{g}=U\Big(dR^2+R^2(\sigma_1^2+\sigma_2^2)\Big)+U^{-1}\sigma_3^2,  \quad \mbox{where}\quad U=\frac{1}{R}-4A
\]
which we recognize as the ASD Taub--NUT metric  written in the Gibbons-Hawking form
\cite{GH}, with
the single-centered harmonic function on $\R^3$ given by $U$.

More precisely, this is the ASD Taub--NUT metric with negative mass. 
In general the K\"ahler form of $g$ given by (\ref{lebrun88}) is $Y=d(r^2\sigma_3)$. If $B=A^2/4$ the 
rescalled two--form 
\[
\hat{Y}:=\Omega^3 Y=64\;R^{3}\; d\Big(\frac{1-4RA}{8R}\sigma_3\Big)
\]
is a conformal Killing--Yano form of 
$\hat{g}=\Omega^2 g$, i. e.
$
\hat{\nabla}_{a}\hat{Y}_{bc}=\hat{\mu}_{abc}+2\hat{g}_{a[b}\hat{K}_{c]}
$
for some three-form $\hat\mu$ and one--form $\hat{K}$. Moreover
$\hat{*}d\hat{Y}$ is the one--form generating the tri-holomorphic $U(1)$ isometry of the
Taub-NUT space\footnote{In \cite{GR} it was shown that the Taub--NUT space also admits a
Killing--Yano tensor $\mathcal{Y}$. This structure is different to the one we have unveiled
as our $\hat{Y}$ is self-dual but $\mathcal{Y}$ does not have a definite duality. 
However there is a relation between both structures as $d\hat{Y}=d \mathcal{Y}$. Therefore
$\mathcal{Y}=\hat{Y}+dB$, where $B$ is a one--form such that $dB$ is a closed conformal Killing--Yano
form.}
\subsubsection*{Fubini--Study metric} The Fubini--Study metric on $\CP^2$ is 
Einsten, ASD and K\"ahler, albeit
with reversed orientation. The ASD Weyl curvature is constant, and therefore
the one--form $V$, and the spinor $W_{A'ABC}$ both vanish, and so 
the scalar invariant (\ref{main_th_1}) vanishes. The tensor invariant (\ref{Tab_tensor}) reduces to the Schouten tensor which is a non--zero constant multiple of the metric.
Thus 
\[
T\neq 0,
\]
and there are no Ricci-flat metrics in the Fubini--Study conformal class.
\subsubsection*{Conformally hyper-K\"ahler homogeneous metric} Consider the left--invariant metric
\[
g={\sigma_0}^2+{\sigma_1}^2+{\sigma_2}^2+{\sigma_3}^2
\]
on a four--dimensional nilpotent Lie group, with the Lie algebra specified by 
the Maurer--Cartan relations
\[
d\sigma_0=2\sigma_0\wedge\sigma_3-\sigma_1\wedge\sigma_2, \quad d\sigma_1=\sigma_1\wedge\sigma_3, 
\quad d\sigma_2=\sigma_2\wedge\sigma_3,\quad
d\sigma_3=0.
\]
Using these relations we find that the Weyl tensor is ASD, but the Ricci tensor is 
non--zero. Calculating the obstructions of Theorem \ref{theo_main_th}  shows that both the scalar and the tensor invariant vanish. Thus the metric is conformal to Ricci-flat. We also find
\[
V=-\frac{3}{2}\sigma_3.
\]
To find the conformal factor introduce the function $z$ on the Lie group such that
$\sigma_3=d\;\mbox{ln}(z)$. The formula (\ref{conf_w}) implies that $V$ can be set to zero 
if the conformal factor $z^3$ is used. Thus the conformally rescaled metric
\be
\hat{g}=z^3\; g
\ee
is ASD and Ricci-flat. We claim that this metric arises from the Gibbons--Hawking 
ansatz \cite{GH}
 with the linear potential. To see this, introduce local coordinates 
$(\tau, x, y, z)$  such that
\[
\sigma_0=z^{-2}(d\tau+ydx), \quad \sigma_1=z^{-1} dx, \quad \sigma_2=z^{-1} dy.
\]
Then the resulting ASD Ricci--flat metric is of the form
\[
\hat{g}=U(dx^2+dy^2+dz^2)+U^{-1}(d\tau+\alpha)^2,
\]
where the harmonic function $U=z$, and the one-form $\alpha=ydx$ satisfy the monopole equation
on $\R^3$
\[
*_3 dU =d\alpha.
\]
\subsubsection*{Conformally K\"ahler ASD example}
Let 
\be
\label{ccasd}
g=dx^2+dy^2+dz^2+\frac{1}{y^2z^2}(d\tau+2xydy-xzdz)^2.
\ee
This ASD metric arises from a particular choice of the Abelian
monopole over the hyperbolic 3-space with its Einstein--Weyl structure
\cite{JT}. The scalar invariant
(\ref{main_th_1}) does not vanish and is given by
\[
\frac{9}{16z^2(z^4+z^2y^2+4y^4)}.
\]
Thus $g$ is not conformal to a Ricci-flat metric. 
The second order obstruction (\ref{Tab_tensor}) also does not vanish. Its anti-symmetric part $T_{[ab]}=\nabla_{[a}V_{b]}$ is 
\[
dV=\frac{3yz^3(z^2+8y^2)}{(z^4+z^2y^2+4y^4)^2}dy\wedge dz.
\]
We note that the metric $g$ is nevertheless conformal to a K\"ahler metric, as can be verified by computing the obstructions of \cite{DT09}.
\subsubsection*{The hyperbolic ansatz and ASD metrics on connected 
sums of $\CP^2$s}
Examples (\ref{lebrun88}) with $B=0$, and (\ref{ccasd}) fall into the class governed by LeBrun's hyperbolic ansatz.  Any scalar--flat K\"ahler metric with $U(1)$ symmetry is locally
of the form
\be
\label{toda_metric}
g={P}(e^u(dx^2+dy^2)+dz^2)+\frac{1}{P}(d \tau +\alpha)^2
\ee
where $u=u(x, y, z)$  is a solution of the $SU(\infty)$ Toda equation
\be
\label{toda}
u_{xx}+u_{yy}+({e^u})_{zz}=0,
\ee
the function $P$ satisfies the linearised  $SU(\infty)$ Toda equation
\be
\label{ltoda}
P_{xx}+P_{yy}+({Pe^u})_{zz}=0,
\ee
and the one--form $\alpha$ satisfies
\be
\label{dalpha}
d\alpha=-P_xdy\wedge dz-P_y dz\wedge dx -(Pe^u)_zdx\wedge dy.
\ee
Here the $U(1)$ action is generated by the Killing vector $K=\p/\p\tau$,
and $(x, y, z)$ are coordinates on the space of orbits of $K$ in $M$,
$z$ is the moment-map generating $K$, and $x+iy$ is the holomorphic
coordinate on the leaf space in $M$ of the foliation by
$JK+iK$, where $J$ is the complex structure on $M$. 
There is one particular solution to equation (\ref{ltoda})
given by $P=u_z$ for which $g$ is Ricci--flat. 

 Consider a solution  to (\ref{toda}) given by $u=2\log{z}$ where $z>0$, and 
introduce the coordinate $q=\sqrt{2z}$. Then  (\ref{toda_metric})
takes the form
\be
\label{le_brun}
g=q^2(Uh+U^{-1}(d\tau+\alpha)^2),
\ee
where 
\[
h=\frac{1}{q^2}\Big(dx^2+dy^2+dq^2\Big)
\]
is the hyperbolic metric, and equation (\ref{ltoda}) implies that
$U=q^2P$ belongs to the kernel of the hyperbolic Laplacian of $h$.
The metric (\ref{lebrun88}) with $B=0$ (called the Burns metric) is of the form (\ref{le_brun}),
where $U$ is the fundamental solution of the hyperbolic Laplace equation
\[
U=1+\frac{1}{\exp{(2\rho)}-1},
\]
and $\rho$ is the hyperbolic distance between the point
$(x, y, q)$ and a fixed point $(0, 0, q_0)$ in the hyperbolic three-space:
\[
\rho=\mbox{cosh}^{-1}\Big(\frac{x^2+y^2+q^2+{q_0}^2}{2qq_0}\Big).
\]
Our analysis leading to (\ref{obstruction_T}) shows that in this case
the metric (\ref{le_brun}) does not contain a Ricci-flat metric in its conformal class.
This metric is conformal to the Fubini--Study metric on $\CP^2$, so in a sense the
non-vanishing of the obstruction $T$ for the Fubini-Study metric
follows from (\ref{obstruction_T}).

In \cite{L91} LeBrun has demonstrated that taking $U$ to be 
a superposition of the fundamental solutions corresponding to $n$ distinct 
points in the hyperbolic space gives rise to 
scalar-flat K\"ahler asymptotically flat metric (\ref{le_brun}) on the blow-up of $\C^2$ at $n$-points along a complex line. Moreover a conformal class
of $g$ contains an ASD metric on a connected sum of $n$ copies of 
$\CP^2$ with reversed orientation. For a given collection of $n$ points $p_1, \dots, p_n\in \HH^3$, let $U_j=(\exp{(2\rho_j)}-1)^{-1}$, where
$\rho_j$ is the hyperbolic distance from $p_j$ to a point with coordinates $(x, y, q)$. Then the conformal class containing the ASD metric on 
a connected sum of $n$ copies of $\CP^2$s is represented by (\ref{le_brun}) with $U=1+\sum_{j=1}^n U_j$.

\begin{theo}
\label{thmlb}
LeBrun's metrics on connected sums of $\CP^2$ are not conformally Ricci-flat on any open set. 
\end{theo}
{\bf Proof}.
 This can be seen by explicitly computing the obstruction $T$ for the given $U$ - it does not vanish, but
the resulting formulae are unilluminating. The MAPLE documentation of the proof is available from the authors on request. 

Instead we give an heuristic argument based on continuity: 
close to any of the points $p_j$ the metric (\ref{le_brun}) can be approximated
by the Burns metric, and (\ref{obstruction_T}) implies that the obstruction
(\ref{Tab_tensor}) is approximately
\[
T\cong-\frac{A}{r^4}\;g.
\]
Thus $g$ does not contain a Ricci-flat metric, at least in the open sets containing the points $p_j$.
\koniec
A related result has been established in \cite{PT}, where it was shown that LeBrun's conformal class
does not contain an Einstein metric with $\Lambda\neq 0$ unless $n=1$, or all points $p_j$ coincide. It is not clear wether the results in
\cite{PT} apply in the limit when $\Lambda=0$.
An anonymous referee has pointed out an elegant alternative
proof of Theorem \ref{thmlb}. His global argument hinges on the fact that LeBrun manifolds are not spin.
\section{Neutral signature and null-K\"ahler structures}
\label{sec22}
If the signature of $g$ is neutral, the Proposition (\ref{prop_riem}) does not hold, and
all sixteen conditions (\ref{BE_inv}) and (\ref{L_minors}) have to vanish if there exist two linearly independent solutions to the twistor equation (\ref{twist_eq}), and $g$ is conformal to a Ricci-flat metric.
In this Section we shall analyze the case where there exists only one real solution to the twistor equation. 
Let us first make the following definition \cite{D02}
\begin{defi}
A null-K\"ahler structure on a $2n$--dimensional manifold $M$  is a pair $(g, N)$
where $g$ is a metric of signature $(n, n)$ and
$N: TM\longrightarrow TM$ is a rank-$n$ endomorphism
such that
\[
N^2=0, \quad g(NX, Y)+g(X, NY)=0, \quad \nabla N=0
\]
for all vector fields  $X, Y$ on $M$. 
\end{defi}
If $n=2$  a null-K\"ahler structure can be equivalently defined by the existence of  
a real spinor $\iota^{A'}$ which is covariantly constant
with respect to the Levi--Civita connection of $g$. Given such a spinor the endomorphism $N$ is
represented by ${N^{a}}_b=\iota^{A'}\iota_{B'}{\epsilon^{A}}_{B}$. 
The isomorphism 
$\mbox{Sym}^2(\spp')\cong\Lambda^2_+$ associates a two-form $\Sigma$ to the spinor $\iota^{A'}$. This form
is explicitly given by $\Sigma(X, Y)=g(NX, Y)$. Therefore $\Sigma$ is simple, i. e. $\Sigma\wedge\Sigma=0$,
and parallel. The following result was established in \cite{D02} under an additional assumption that
$g$ is ASD. Here we give a general proof.
\begin{prop}
A neutral signature metric $g$ is conformal to a null-K\"ahler metric
if and only if there exist a real spinor $\iota_{A'}$ which satisfies the twistor equation
\be
\label{twist_eq_22}
\nabla_{A(A'}\iota_{B')}=0.
\ee
\end{prop}
{\bf Proof}.
Dropping the symmetrisation yields
\be
\label{eq111}
\nabla_{AA'}\iota_{B'}=\epsilon_{A'B'}\alpha_A
\ee
for some spinor $\alpha_A$ as in (\ref{t1}). We aim to construct a conformal 
factor, but the ideas behind the proof of Proposition \ref{prop_twist_eq} 
have to be modified as the formula (\ref{conf_factor}) does not make sense if there is only one real solution to the twistor equation. First note that 
(\ref{eq111}) implies 
\[
\iota^{A'}\iota^{B'}\nabla_{AA'}\iota_{B'}=0
\]
so that the rank--two 
distribution
 $<\delta_0, \delta_1>= \mbox{Ker}(\Sigma)\subset TM$
is integrable. We claim that this is 
sufficient to satisfy the integrability 
conditions for the existence of a function $\phi$ such that
$\alpha_{A}=\delta_A \phi$. Indeed, consider 
the overdetermined system
$\alpha_0=\delta_0\phi, \alpha_1=\delta_1\phi$. The Frobenius theorem 
gives
\be
\delta_0\alpha_1-\delta_1\alpha_0=\beta^A\alpha_A
\label{ba}
\ee
for some $\beta^A$.
On the other hand from (\ref{eq111}) ${\nabla^A}_{A'}\alpha_A=2\Lambda \iota_{A'}$ which  implies
$\iota^{A'}{\nabla_{A'}}^A\alpha_{A}=0$. Therefore  $\delta^A\alpha_A=\iota^{A'}{\Gamma_{AA'}}^{AC}\alpha_C$
which is (\ref{ba}) with $\beta^C=\iota^{A'}{\Gamma_{AA'}}^{AC}$. From (\ref{conf_change_pi}) it follows
that under conformal rescallings $\hat{\alpha}_A=\alpha_A+\Omega^{-1}\iota^{A'}\nabla_{AA'}\Omega$.
Thus we chose the conformal factor $\Omega=\exp{(-\phi)}$ to set $\hat{\alpha}_A=0$.
\koniec 
\subsection{Null K\"ahler structures conformal to Einstein}
It is known that in four dimension an ASD Einstein metric with $\Lambda\neq 0$ is conformally equivalent to a K\"ahler metric if and only if
it admits an isometry \cite{DT09}. We shall establish an analogue of this result in the null-K\"ahler context.
\begin{prop}
\label{NK_prop}
Let $g$ be an Einstein metric with a non--zero cosmological constant which is conformally equivalent to a null-K\"ahler metric. Then $g$ admits a hyper-surface-orthogonal null Killing vector $K$ such that the self--dual derivative of $K$ is proportional to the null--K\"ahler two--form $\Sigma$.
Morevover
the conformal factor relating the Einstein and the null-K\"ahler metrics is constant along this Killing vector.
\end{prop}
{\bf Proof.} We shall work in the Einstein scale, where $\Phi_{ABA'B'}=0$
and (\ref{t2}) yields
\[
\nabla_a \iota_{B'}=\epsilon_{A'B'}\alpha_A, \quad \nabla_a\alpha_{B}=\Lambda\epsilon_{AB} \iota_{A'}
\]
where $\alpha^A\neq 0$, 
and $\psi_{ABCD}\alpha^A=0$ so that  $\psi_{ABCD}$ is of type $N$. We also find that
\[
K_a=\alpha_{A}\iota_{A'}
\]
is a null Killing vector
because $|K|^2=(\alpha_A\alpha^A)(\iota_{A'}\iota^{A'})=0$ and
\[
\nabla_a K_b=\Lambda \epsilon_{AB}\iota_{A'}\iota_{B'}+\epsilon_{A'B'}\alpha_A \alpha_B,
\]
so the RHS is a two--form.
This Killing vector is twist-free (i. e. it is orthogonal to a hyper-surface)  as
\[
K_{a}\nabla_b K_{c}=\Lambda \epsilon_{BC}\iota_{B'}\iota_{C'}\iota_{A'} \alpha_A
+\epsilon_{B'C'}\alpha_A \alpha_B \alpha_C \iota_{A'}
\]
and $(\star(K\wedge dK))_a={\epsilon_a}^{bcd}K_b\nabla_c K_d=0$ 
as $\epsilon^{abcd}$ involves contraction over a pair of primed and unprimed indices.

Finally we shall demonstrate that the conformal factor in the ASD Einstein metric is Lie derived along the null Killing vector, and thus the null-K\"ahler metric also admits a null Killing vector (rather than a conformal null Killing vector). Indeed,
if $\alpha_A=\iota^{A'}\nabla_{AA'}\phi$, where $\exp{(-2\phi)}$ is the conformal factor then
\[
{\mathcal L}_K(\phi)=\iota^{A'}\alpha^A\nabla_{AA'}\phi=\alpha^A\alpha_A=0.
\]
\koniec
\subsubsection{ASD cosmological plane waves}
We shall now construct a class of examples of ASD Einstein metrics
conformal to null-K\"ahler metrics.
Any null-K\"ahler metric is locally given by
\be
\label{nk_form_metric}
g=dwdz+dydz-\Theta_{xx}dz^2-\Theta_{yy}dw^2+2\Theta_{xy}dwdz,
\ee
where $\Theta=\Theta(x, y, z, w)$ is an arbitrary function of four variables
with continuous 4th derivatives \cite{D02}. The null-K\"ahler structure is given by 
a parallel simple two-form
$\Sigma=dw\wedge dz$. The Weyl tensor is anti--self--dual if $\Theta$ satisfies a 4th order PDE
\begin{eqnarray}
\label{NK_equations}
&&f:=\Theta_{wx}+\Theta_{zy}+\Theta_{xx}\Theta_{yy}-\Theta_{xy}^2,\nonumber\\
&&f_{wx}+f_{zy}+\Theta_{yy}f_{xx}+\Theta_{xx}f_{yy}-2\Theta_{xy}f_{xy}=0.
\end{eqnarray}
Now impose the HSO null Killing vector condition from Proposition (\ref{NK_prop})
\[
{\mathcal L}_K g=0, \quad g(K, K)=0, \quad {\mathcal L}_K(dw \wedge dz)=0, \quad K\wedge dK=0.
\]
One example
of such null Killing vector is given by $K=\p/\p {y}$. 
The most general null-K\"ahler metric which admits this Killing vector is
conformally equivalent to 
\[
\hat{g}=\frac{2}{{x}^2}(dwd{x}+dz d{y}+\Lambda dw^2-Fdz^2).
\]
It is a rescaling, by a factor of $x^{-2}$ of the metric
(\ref{nk_form_metric}) with 
\[
\Theta_{{y}{y}}=-\Lambda, \quad \Theta_{{x}{y}}=0,
\quad \Theta_{{x}{x}}=F,
\]
where $F=F(x, w, z)$. The conformal factor has been chosen to make
the metric $\hat{g}$ Einstein with cosmological constant given by $\Lambda$.
Imposing the anti--self--duality condition   (\ref{NK_equations}) on the Weyl spinor gives 
\[
F=f({x}+\Lambda w, z)-\frac{1}{2}{x}\dot{f} ({x}+\Lambda w, z)
\]
where dot denotes the differentiation w.r.t the first argument.
Change coordinates ${y}=\tilde{y}, {x}=\tilde{x}-\Lambda w$
so that\footnote{We have been informed by Adam Chudecki and Maciej Przanowski 
that (\ref{generalised_H}) is the most general local form of ASD Einstein metric with $\Lambda\neq 0$ which admits a null Killing vector \cite{CP13}. This provides a converse to Proposition
to (\ref{NK_prop}):
An  ASD Einstein metric is conformal  to a null-K\"ahler metric if and only if it admits a null
Killing vector.} (dropping tildes)
\be
\label{generalised_H}
\hat{g}=\frac{2}{(x-\Lambda w)^2}\Big(dwdx+dzdy-\Big(f(x, z)-\frac{1}{2}(x-\Lambda w) \p_x{f}(x, z)\Big) dz^2\Big).
\ee
In the limit $\Lambda=0$ we obtain an ASD Ricci-flat plane wave.
If the function $f$ in  (\ref{generalised_H}) does not depend on the coordinate $z$, then $g$ admits
a non--null Killing vector $\p/\p z$ in addition to the null Killing vector. The metric $\hat{g}$ 
corresponding to this
case has been found by Hoegner \cite{Hoegner}. Hoegner has shown that the Einstein--Weyl structure
\cite{JT} on the space of orbits of $\p/\p z$ in $\hat{g}$ is the most general Einstein--Weyl structure
which is simultaneously of dKP  and $SU(\infty)$ Toda type.
\subsubsection{Connection with projective structures}
The null Killing vector from Proposition \ref{NK_prop}
defines a  pair of
totally null foliations of $M$, one by $\alpha$-surfaces and one by
$\beta$-surfaces. These foliations intersect along integral curves of
$K$ which are null geodesics.
The $\beta$-plane
distribution is spanned by $\alpha^A$, and it follows from the Killing equation
that it is  integrable. In \cite{DW07} it was shown that there exists a canonically defined
projective structure on 
the two-dimensional space of
$\beta$-surfaces $U$ which arises as a quotient of ${M}$ by the
$\beta$--plane
rank--two 
distribution $\mbox{Ker}(\Omega)\subset TM$. Here $\Omega_{ab}=\alpha_A\alpha_B\epsilon_{A'B'}$ is the ASD two--form corresponding 
to the spinor $\alpha\in\Gamma(\spp)$ under the isomorphism $\Lambda^2_-(M)\cong
\spp\odot\spp$. Recall that a projective structure is an
equivalence class of connections, where two connections are
equivalent if they have the same unparameterized geodesics. In two dimensions
projective structures are locally the same as second order ODEs of the form
\begin{equation} \label{secondorderode}
\frac{d^2 Y}{dX^2} = A_3(X,Y) \Big( \frac{d Y}{dX} \Big)^3 +
A_2(X,Y) \Big( \frac{dY}{dX} \Big)^2 + A_1(X,Y) \Big( \frac{dY}{dX}
\Big) + A_0(X,Y).
\end{equation}
The ODE is obtained by selecting local coordinates $(X,Y)$ on $U$ and eliminating the
affine parameter from the geodesic equation. The functions $A_i(X, Y)$ can be
expressed in terms of combinations of connection coefficients. 

The normal form
of the ASD conformal structure on $M$ with a null Killing vector depends on whether
the null vector is twisting or not. In the non--twisting case (which is relevant here)
it is given by
\begin{eqnarray}
\label{non_twist}
g &=& \Big(d T + (Z A_3 - Q) dY\Big)\Big(dY - \beta dX\Big) - \Big(dZ - (Z (-\beta_Y +
A_1 + \beta A_2 + \beta^2 A_3))dX\nonumber \\
& &- (Z(A_2 + 2 \beta A_3)+ P)dY\Big)dX,
\end{eqnarray}
where $(X, Y, T, Z)$ are local coordinates on $M$ such that the null 
Killing vector is $\p/\p T$. 
The arbitrary functions
 $A_1,A_2,A_3,\beta,Q,P$ depend on  $(X,Y)$ and
the function
$A_0$ is given by
\[
A_0 = \beta_X + \beta \beta_Y - \beta A_1 - \beta^2 A_2 - \beta^3 A_3.
\]
To read off the projective structure from (\ref{generalised_H}) we disregard the  
the conformal factor (the projective structure depends only on the conformal structure)
and set
\[
T=y, \quad Z=-w, \quad Y=z, \quad X=x.
\]
The resulting metric is of the form (\ref{non_twist}) with  $A_1=A_2=\beta=P=0$.
The 2nd order ODE
defining this projective structure is
\[
\frac{d^2 Y}{d X^2}=\frac{\Lambda}{2}\p_X{f}(X, Y) \Big(\frac{d Y}{d X}\Big)^3.
\]
The projective curvature \cite{BEG,BDE} is 
\[
\frac{\Lambda}{2}\frac{\p^3 f}{\p X^3}dY\otimes (dX\wedge dY).
\]

\subsection{Outlook}
We have constructed a rank four vector bundle $E\rightarrow M$ with connection ${\mathcal D}$ such that the parallel  sections of this connection correspond to Ricci-flat metrics in an anti-self-dual conformal class. This has lead to a complete set of obstructions to the existence
of a Ricci-flat metric in a given ASD conformal class (Theorem \ref{theo_main_th}).

The curvature of the connection ${\mathcal D}$ is ASD (Proposition \ref{prop_asd_1}) , so it can be regarded as an ASD Yang--Mills field. Thus, by the Ward transform \cite{ward} (or its generalisation to backgrounds with ASD conformal curvature), $E$ corresponds to some holomorphic vector bundle   over the twistor space
which is holomorphically trivial on the twistor curves. This holomorphic vector bundle
can be also constructed directly from the twistor data
following the related constructions of LeBrun \cite{L86} and Merkulov \cite{merkulov}. In $(2, 2)$ signature there is an alternative approach \cite{Gro},  where the twistor curves arise as integral curves of systems of second order ODEs with vanishing Wilczynski invariants. The tensor obstructions 
from Theorem \ref{theo_main_th} should have their counterparts as point invariants of the corresponding system of ODEs.
Some invariants of these type under a more restrictive fibre preserving transformations
have been found in \cite{CDT12}. An outstanding, and interesting open problem is to generalise Theorem \ref{theo_main_th} to the case of ASD Einstein metrics with non--zero scalar curvature.


\end{document}